\documentclass[11pt, a4paper]{article}
\usepackage{jcappub}

\usepackage[utf8]{inputenc} 
\usepackage[T1]{fontenc}    
\usepackage{lmodern}
\usepackage{hyperref}       
\usepackage{url}            
\usepackage{amsfonts}       
\usepackage{amsmath,amsthm,amssymb} 
\usepackage{bm}
\usepackage{bbm}
\usepackage{dcolumn}
\usepackage{nicefrac}       
\usepackage{graphicx}
\usepackage{xcolor, etoolbox}
\usepackage{mathtools}
\usepackage{mathrsfs}
\usepackage{soul}
\usepackage{ragged2e}
\usepackage{microtype}      
\usepackage{braket}      
\usepackage{enumitem}
\usepackage{xspace}
\usepackage{dsfont}
\usepackage{float}
\usepackage[export]{adjustbox}

\DeclarePairedDelimiter\abs{\lvert}{\rvert}
\renewcommand{\vec}[1]{\boldsymbol{#1}}

\hypersetup{
    colorlinks,
    linkcolor=blue,
    citecolor=blue,
    urlcolor=blue,
    breaklinks=true 
}

\graphicspath{{./figures/}} 

\begin{document}

\title{\boldmath Late vacuum choice and slow roll approximation in gravitational particle production during reheating}

\author[a]{Jose A. R. Cembranos,}
\author[a]{Luis J. Garay,}
\author[a]{Álvaro Parra-López}
\author[b]{and Jose M. Sánchez Velázquez}

\affiliation[a]{Departamento de F\'isica Te\'orica and IPARCOS, Facultad de Ciencias F\'isicas, \\
Universidad Complutense de Madrid,  Plaza de las Ciencias 1, 28040 Madrid, Spain} 
\affiliation[b]{Instituto de F\'isica Te\'orica UAM/CSIC, c/ Nicol\'as Cabrera 13-15,\\ Cantoblanco, 28049, Madrid, Spain}

\emailAdd{cembra@ucm.es}
\emailAdd{luisj.garay@ucm.es}
\emailAdd{alvaparr@ucm.es}
\emailAdd{jm.sanchez.velazquez@csic.es}
	
\date{\today}

\abstract{In the transition between inflation and reheating, the curvature scalar typically undergoes oscillations which have significant impact on the density of gravitationally produced particles. The commonly used adiabatic vacuum prescription for the extraction of produced particle spectra becomes a non-reliable definition of vacuum in the regimes for which this oscillatory behavior is important. In this work, we study particle production for a scalar field non-minimally coupled to gravity, taking into account the complete dynamics of spacetime during inflation and reheating. We derive an approximation for the solution to the mode equation during the slow-roll of the inflaton and analyze the importance of Ricci scalar oscillations in the resulting spectra. Additionally, we propose a prescription for the vacuum that allows to safely extrapolate the result to the present, given that the test field interacts only gravitationally. Lastly, we calculate the abundance of dark matter this mechanism yields and compare it to observations.}

\keywords{Cosmology of Theories beyond the SM, Effective Field Theories, Classical Theories of Gravity}

\arxivnumber{2301.04674}
~\hfill IPARCOS-UCM-23-002\par
~\hfill IFT-UAM/CSIC-23-4\par
\maketitle
\flushbottom


\section{Introduction}
\label{sec:introduction}

The theory of quantum fields in curved spacetimes accommodates a plethora of unexpected phenomena such as Hawking radiation \cite{Hawking1975}, the Unruh effect \cite{Unruh1981}, or entanglement across horizons \cite{Gibbons1977, Bombelli1986, Page1993, Srednicki1993}, that have changed our perspective on the interplay between quantum fields and gravity. Gravitational particle production due to the spacetime dynamics \cite{Parker1969, Ford1987} is one of these phenomena and can be particularly important during the early stages of the universe, since it may be able to explain the dark matter abundance, as it has been extensively discussed in the literature \cite{Chung1998,Chung2001,Chung2019, Hashiba2019, Ema2016,Ema2018,Markkanen2017a,Cembranos2020,Ema2019,Bastero2019,Markkanen2018, Herring2020,Fairbairn2019,Kainulainen2023,Ford2021}. The rapidly evolving spacetime during inflation \cite{Starobinsky1980, Guth1981, Linde1982} and the consequent transient to reheating \cite{Kolb1990, Kofman1994, Kofman1997, Allahverdi2010, Baumann2015} can produce a significant density of particles for any field non-conformally coupled to the geometry, regardless of its interaction with other fields. Therefore, it is of particular interest to analyze this phenomenon from the point of view of a dark matter production mechanism. In the absence of interactions, the abundance of dark matter produced in the early universe due to the expansion of spacetime is not diluted as a consequence of thermalization with other fields. It remains then as a relic abundance, so that this mechanism alone can in fact explain current observations. This has been mostly explored for scalar fields that are non-minimally coupled to gravity in many works. In particular, in refs. \cite{Chung1998, Chung2001, Chung2019, Hashiba2019}, the authors study the production of supermassive dark matter candidates (WIMPZillas), and, more recently, references \cite{Ema2016, Ema2018, Markkanen2017a, Cembranos2020} incorporated the importance of the oscillatory behavior of the background geometry for the production. On the other hand, gravitational production of more general fields, such as fermion and vector fields, has also been analyzed in \cite{Ema2019, Bastero2019}. Usually, the dark matter candidate is regarded as a \emph{spectator} field~\cite{Markkanen2018, Herring2020} which does not source gravity, and with no direct coupling to the inflationary fields. However, it is generally non-minimally coupled to the geometry via the curvature scalar, and interactions with other fields are disregarded. In all these works, it is customary to make use of the adiabatic prescription to define the vacuum state of the dark matter field in order to calculate the gravitational production. This definition seems to hold after a few oscillations of the inflaton in the reheating stage, but only in the case of very large masses of the dark matter candidate. In the regime of low masses, however, this vacuum provides a correct prediction only when considering very late times, after many oscillations have occured. Importantly, this oscillating behavior influences gravitational production \cite{Cembranos2020}. It is worth mentioning that the type of dark matter produced in this way is adiabatic \cite{Markkanen2017a, Tenkanen2019}, and therefore the observational constraints on isocurvature perturbations \cite{Planck2018} do not have to be considered.

In this work, we study the gravitational production of a massive scalar field $\varphi$ described by a Klein-Gordon action that includes a non-minimal coupling to the Ricci curvature scalar~$R$ through a term of the form~$\xi R\varphi^2$. The strength of this coupling is determined by the parameter~$\xi$. In an attempt to accommodate the arguments put forward in refs. \cite{Markkanen2017b, Fairbairn2019, Wang2019, Cembranos2020} concerning vacuum instability, overproduction, and quantum cosmology analyses, we restrict ourselves to the range $1/6 \leq \xi \leq 1$ for the coupling constant $\xi$. More explicitly, values of~$\xi$ smaller than $1/6$ would lead to tachyonic instabilities in the asymptotic past. On the other hand, although for a general scalar field it is possible to have $\xi>1$ (contrarily to what happens in the case of ref. \cite{Markkanen2017b}), the behavior of particle production is in this case qualitatively the same as for $\xi \sim 1$. Nevertheless, interesting results have been obtained for the regime of very large $\xi$ when the back-reaction of the created field is taken into account~\cite{Lebedev2022}. We have considered an inflationary epoch driven by a single inflaton field~$\phi$ that slowly rolls down a quadratic potential and starts oscillating around its minimum, leading then to a reheating phase. Although this particular model, in its simplest version, is ruled out by CMB observations, it allows us to compare our results with previous literature (see e.g.~\cite{Markkanen2017a, Cembranos2020, Markkanen2018, Fairbairn2019, Kainulainen2023}), and, furthermore, our analysis can be carried out in the same way for more realistic potentials. The dynamics for the inflaton is analytically solved at the onset of inflation, while the transition to the reheating epoch is modeled numerically. Our scalar field is assumed to be in the Bunch-Davies vacuum state when inflation starts. In order to extract the gravitational production, the Klein-Gordon equation of the field $\varphi$ is solved from that point in time until the dynamics enters the adiabatic regime and particle production becomes negligible. Moreover, one also needs to provide a definition of vacuum for this instant, for which the adiabatic prescription is usually adopted. We discuss its validity and introduce as well an \emph{averaged} vacuum that produces the same density of particles but allows to obtain the correct result much earlier than the time at which adiabaticity is reached. This is particularly helpful when considering masses way below the inflaton mass for our scalar field, which are of great interest concerning dark matter candidates. Also, we stress the importance of taking into account the first few hundreds of oscillations of the inflaton in the final prediction and present the results in the form of spectra and total density of produced particles for different values of the scalar field mass $m$ and its coupling $\xi$ to the Ricci scalar.   

For obtaining the gravitationally produced particle density, we develop an analytic solution to the mode equation that is valid during slow-roll. As opposed to a full numerical solution, this increases the efficiency of evaluations while allows us to qualitatively understand the behavior of the field during this process. In particular, by making a controlled analysis of the errors, this analytic solution allows for a study of the impact of assuming slow-roll during the whole inflationary period. Regardless of the method used for obtaining particle production, this manuscript focuses on the choice of late vacuum and the impact of curvature oscillations in the final spectra.

The remainder of this paper is organized as follows. In section \ref{sec:scalarfield}, we introduce the field that is coupled to the expanding geometry, and work out the formalities of Bogoliubov-like particle production in this context. In order to determine the complete form of the mode equation, we need to provide the background dynamics coming from the particular inflationary model in consideration, which we do in section \ref{sec:background}. With all these ingredients, we explore the gravitational production for the scalar field in section~\ref{sec:particleproduction}, analyzing the solution to the mode equation in the different regimes and studying the influence of the oscillations of the curvature scalar in the final result. Moreover, we discuss the importance of the vacuum choice when obtaining the number density of produced particles. Lastly, we present our results in the form of spectra and total density of particles in section~\ref{sec:spectra} and elaborate our conclusions in section \ref{sec:conslusions}.   

\emph{Notation.}  We set $M_{\text{P}}=1/\sqrt{G}, \hbar = c = k_{\text{B}} = 1$, and use the metric signature $\left(-, +, +, +\right)$. Furthermore, greek indices $\mu, \nu$ run from $0$ to $3$, while latin indices $i,j$ run from $1$ to $3$.

\section{Dynamics of a scalar field in flat FLRW cosmologies}
\label{sec:scalarfield}

We will consider a massive scalar field $\varphi$ non-minimally coupled to gravity in a Friedmann-Lemaître-Robertson-Walker (FLRW) spacetime with vanishing spatial curvature \cite{Friedman1922, Friedman1924, Lemaitre1931, Robertson1935, Robertson1936a, Robertson1936b, Walker1937}. We will not consider here any coupling of the derivatives of the scalar field (see \cite{Borrajo2020}). 

The dynamics of our scalar field is encoded in the action
\begin{equation}
S = -\frac{1}{2}\int d ^4x \sqrt{-g}\left[\partial_{\mu}\varphi \partial^{\mu}\varphi + \left(m^2 + \xi R\right)\varphi^2\right],
\label{eq:GeneralAction}
\end{equation}
where $g$ is the determinant of the metric, $m$ is the bare mass of the field, and $\xi$ is the coupling   to the Ricci curvature scalar $R$. As it is well-known, this form of interaction is required to provide a renormalizable theory of scalar field.
The geometry is determined by the  spatially flat FLRW line element
\begin{equation}
d s^2 = a^2(\eta)\left(-d \eta^2+d x^2 + d y^2 + d z^2\right),
 \label{eq:GeneralLineElement}
\end{equation}
where we have considered Cartesian coordinates for the flat spatial sections, and $\eta$ is the conformal time, related to cosmological time by $a(\eta)d\eta=d t$. 

It is convenient to work with the auxiliary field  
\begin{equation}
\chi(\eta, \vec{x}) = a(\eta) \varphi(\eta, \vec{x}),
\end{equation}
whose equation of motion can be obtained from the action \eqref{eq:GeneralAction},
\begin{equation}
\chi^{\prime\prime}(\eta, \vec{x}) - \left\{\Delta - a^2(\eta)\left[m^2 + \left(\xi-1/6\right) R\right]\right\}\chi(\eta, \vec{x})=0,
\label{eq:EOM}
\end{equation}
where $\Delta$ is the Laplace operator, $R = 6a^{\prime\prime}/a^3$, and the prime denotes derivative with respect to conformal time.

We can use the eigenfunctions of the Laplace operator, which in our case are Fourier modes, as a basis of functions to expand the scalar field $\chi$,
\begin{equation}
\chi(\eta, \vec{x}) = \int \frac{d^3\vec{k}}{\left(2\pi\right)^{3/2}}\left[a_{\vec{k}}v_k(\eta)
+ a_{-\vec{k}}^*v_k^*(\eta)\right]e^{i\vec{k}\vec{x}},
\label{eq:FieldExpansion}
\end{equation}
where the coefficients $a_{\vec{k}}, a_{\vec{k}}^*$ become creation and annihilation operators upon quantization of the field, with the standard commutation relations \cite{Birrell1982, Mukhanov2007, Calzetta2008, Parker2009}. The time-dependent mode functions $v_k(\eta)$ and $v_k^*(\eta)$ satisfy a harmonic oscillator equation
\begin{equation}
v^{\prime\prime}_k(\eta) + \omega_k^2(\eta) v_k(\eta) = 0,
\label{eq:AuxiliaryModeEquation}
\end{equation}
with $k = \sqrt{\vec{k}^2}$ and a time-dependent frequency
\begin{equation}
\omega_k^2(\eta) =k^2 + a^2(\eta)\left[m^2 + (\xi-1/6) R(\eta)\right].
\label{eq:Frequency}
\end{equation}
The solutions to \eqref{eq:AuxiliaryModeEquation} have to fulfill the normalization condition
\begin{equation}
v_kv^{\prime\, *}_k - v^{\prime}_k v_k^* = i,
\label{eq:WronskianModes}
\end{equation}
so that they are compatible with the standard commutation relations of creation and annihilation operators.

For a given evolution of the background geometry, encoded in the scale factor $a(\eta)$ and the Ricci scalar $R(\eta)$, both \eqref{eq:WronskianModes} and \eqref{eq:AuxiliaryModeEquation} are sufficient to determine $v_k(\eta), v_k^*(\eta)$, which is a basis of the space of solutions of the mode equations. Since any other solution can be expressed as a linear combination of $v_k(\eta)$ and $v_k^*(\eta)$, any two sets of solutions $v_k(\eta)$ and~$u_k(\eta)$ must be related by $u_k = \alpha_k v_k + \beta_k v_k^*$, where normalization \eqref{eq:WronskianModes} on the temporal modes implies the relation $\abs{\alpha_k}^2 - \abs{\beta_k}^2 = 1$ for the complex coefficients $\alpha_k$ and $\beta_k$, which are known as Bogoliubov coefficients \cite{Birrell1982}. Note that the expansion \eqref{eq:FieldExpansion} can be carried out using either basis of solutions. 

Upon quantization of the field, both sets of coefficients $a_{\vec{k}}$ and $b_{\vec{k}}$ (associated with the basis $v_k$ and $u_k$, respectively) and their complex conjugates become operators that give rise to two different definitions on quanta and vacua \cite{Mukhanov2007},
\begin{equation}
\hat{a}_{\vec{k}}\ket{0^{a}} = 0 \quad \text{and} \quad \hat{b}_{\vec{k}}\ket{0^b} = 0, \quad \forall\, \vec{k}.
\end{equation}
These two quantizations are related by the Bogoliubov transformation $\hat{b}_{\vec{k}} = \alpha_k^* \hat{a}_{\vec{k}} - \beta_k^* \hat{a}_{\vec{k}}^{\dagger}$.

The mean number density of $b$-particles in the $a$-vacuum, which will be, in general, a non-vacuum state according to the $\hat{b}_{\vec{k}}$ operators, is given by
\begin{equation}
\bra{0^a}\hat{n}_k^b\ket{0^a} = \abs{\beta_k}^2.
\end{equation}
Integrating over all modes, we find the total mean density $\int d^3\vec{k}\,\abs{\beta_k}^2$, which will remain finite as long as $\abs{\beta_k}^2 \to 0$ faster than $k^{-3}$ for increasing $k$.

Let us now associate each basis of solutions to two observers living at different times~$t_a<t_b$. If spacetime is static, the frequency \eqref{eq:Frequency} is constant, so that the solution to~\eqref{eq:AuxiliaryModeEquation} takes the same form at all times.  As a consequence, observers at different times have the same notion of particle, and therefore $\beta_k = 0$. However, if geometry undergoes an expansion, two observers living at different times (before and after the expansion) have different notions of vacuum. Thus, $\beta_k \neq 0$ and therefore $n^b \neq 0$, which can be understood as the number density of particles produced out of the original vacuum state due to the expansion of spacetime. 

For the problem at hand, the goal is to extract the number of produced particles after the evolution of the universe during inflation and reheating, once these stages have finished. Then, as long as the test particle is not  (strongly)  interacting, this will be related to the abundance one observer would measure today only by the expansion dilution. Hence, we will take the Bunch-Davies vacuum as initial state, as defined by the solution of the mode equation at very early times. In our case, we will take the geometry to approach de Sitter spacetime at the beginning of inflation. On the other hand, the notion of vacuum for an inertial observer after reheating will be different. If the evolution of spacetime is sufficiently adiabatic after this phase, we can assume this is the same vacuum we observe nowadays. Therefore, the corresponding operators will measure the number of particles  created in the evolution.

The specific form of the scale factor and the Ricci scalar will be determined by the specific inflationary model under consideration, which we describe in the next section.

\section{Background dynamics}
\label{sec:background}

We will describe the early epoch of the universe with a chaotic inflationary model consisting of a single scalar field $\phi$ with a quadratic potential of the form $V(\phi) = \frac{1}{2}m_{\phi}^2\phi^2$, where $m_{\phi}$ denotes the inflaton mass. The equation of motion for the inflaton is, if we assume homogeneity and isotropy, 
\begin{equation}
0 = \ddot{\phi} + 3 H(t)\dot{\phi} + \partial_{\phi}V(\phi),
\label{eq:InflatonEOMFlat}
\end{equation}
where $H(t) \equiv \dot{a}(t)/a(t)$ is the Hubble parameter. Note that in this context it is customary to work with cosmological time $t$. 
We will assume that the inflaton contribution to the total energy-momentum tensor is dominant when deriving the corresponding Friedmann equation,
\begin{equation}
H^2 = \frac{4\pi}{3M_P^2}\left[\dot{\phi}^2 + 2V(\phi)\right].
\label{eq:HubbleRate}
\end{equation}
We will also need the Ricci curvature scalar in order to properly describe the frequency of the mode equation \eqref{eq:AuxiliaryModeEquation}, which in terms of the inflaton field reads
\begin{equation}
R =\frac{8\pi}{M_P^2}\left[4V(\phi) - \dot{\phi}^2\right].
\label{eq:RicciStressEnergy}
\end{equation}

\begin{figure}[t!]
\includegraphics[width=0.996\textwidth]{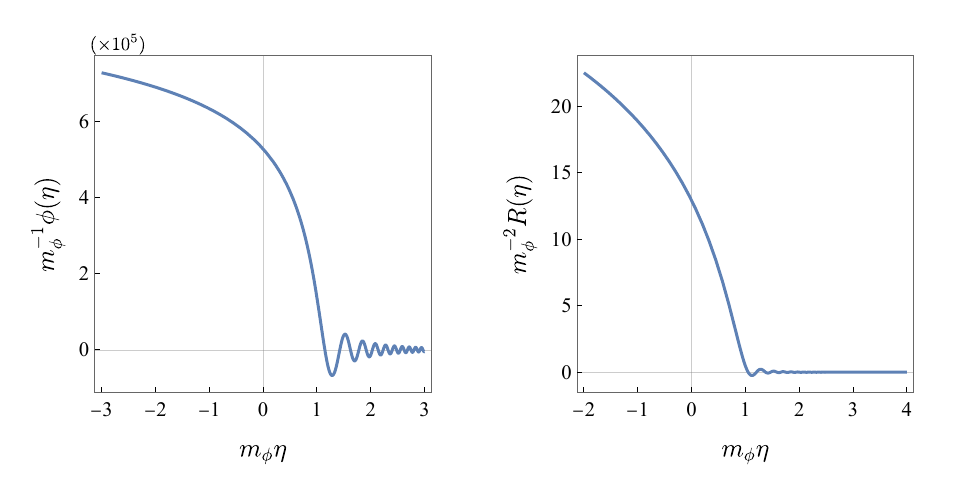}
\begin{picture}(0,0)
\put(316, 124){\includegraphics[width=0.205\textwidth]{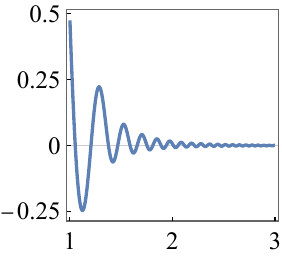}}
\end{picture}
\caption{Inflaton field $\phi(\eta)$ (left panel) and curvature scalar $R(\eta)$ (right panel) as functions of conformal time. The range of time corresponds to the end of inflation and the beginning of reheating. The parameters used for all figures in this article are given in Appendix \ref{app:parameters}.}
\label{fig:InflatonAndRicciConformal}
\end{figure}

Equation \eqref{eq:InflatonEOMFlat}, together with \eqref{eq:HubbleRate}, has no analytic solution in general. However, one can find approximations for certain regimes. When this is not possible, we must rely on numerical computation. We analyze two different regions which, in conformal time, correspond to
\begin{equation}
\eta = \begin{cases}
\eta_{\text{i}} \leq \eta < \eta_*, \quad \text{ Slow-roll approximation},\\
\eta_* \leq \eta \leq \eta_{\text{f}}, \quad \text{Numerical solution}.
\end{cases}
\end{equation}
For the inflationary period, we can use the well-known slow-roll approximation to obtain a solution to the inflaton equation of motion, as we describe in subsection \eqref{subsec:SlowRoll}. However, during the transition between inflation and reheating, the dynamics of the inflaton has to be obtained numerically. Both the inflaton field $\phi$ and the Ricci scalar $R$ start to oscillate with decreasing amplitude, as can be observed in figure~\ref{fig:InflatonAndRicciConformal}, where $\phi(\eta)$ and $R(\eta)$ are depicted for an interval of time during the transition phase. This is the epoch in which most of the particles are produced and the inflaton dynamics is solved until a numerically accessible time~$\eta_{\text{f}}$ is reached, when production becomes negligible. For late times, deep in the reheating era, we can also use an analytic approximation for the solution of the inflaton equation of motion, given in subsection \ref{subsec:latereheating}, which --- although not used in our calculations --- will be used to make some remarks in section \ref{sec:particleproduction}.

\subsection{Inflationary era - Slow-roll approximation}
\label{subsec:SlowRoll}

We will  choose the inflationary period to start at the negative, initial time $t_i$. Inflation requires that the inflaton field changes slowly in comparison to the potential. Within the slow-roll approximation \cite{Mukhanov2005, Weinberg2008}, we can neglect the derivative of the field in favor of the potential, namely $\dot{\phi}^2 \ll \abs{V(\phi)}$. When this condition is satisfied, the field slowly rolls over until it falls to a minimum and starts oscillating. At this point, inflation ends. With this assumption, we can approximately write \eqref{eq:HubbleRate} during the slow roll as
\begin{equation}
H\simeq \sqrt{\frac{8\pi}{3M_P^2}V(\phi)}.
\end{equation}
A slowly-varying inflaton implies that $H \sim \text{constant}$ for this regime. Hence, the expansion of spacetime is said to be  {quasi}-exponential, as it resembles the pure de Sitter solution. Usually, one also assumes a small rate of change for the (already slow) velocity of $\phi$, such that $\abs{\ddot{\phi}} \ll 3H\abs{\dot{\phi}}$. This allows the slow-roll condition to be maintained long enough to solve the flatness and horizon problems. With these assumptions, equation \eqref{eq:InflatonEOMFlat} becomes easily solvable,
\begin{equation}
\dot{\phi}\simeq -\frac{\partial_{\phi}V(\phi)}{3H}\simeq -\partial_{\phi}V(\phi)\frac{M_P}{\sqrt{24\pi V(\phi)}}.
\label{eq:SlowRollInflatonEOM}
\end{equation}
For the particular potential $V(\phi)=\frac{1}{2}m_{\phi}^2\phi^2$, the solution to \eqref{eq:SlowRollInflatonEOM} is
\begin{equation}
\phi_{\text{SR}}(t) = \phi_0 - \frac{M_P}{\sqrt{12\pi}}m_{\phi} t,
\end{equation}
where $t<0$ corresponds to the inflationary period. Note that $t=0$ and $\phi_0$ are the ending time of inflation and the value of the field at this instant, respectively. From here, it is straightforward to obtain an explicit expression for the Ricci scalar, introducing the solution into \eqref{eq:RicciStressEnergy}.

The scale factor is obtained by integrating the Hubble rate, and in the slow-roll approximation it reads
\begin{equation}
a_{\text{SR}}(t) \simeq a_0 e^{-\int_{\phi_0}^{\phi(t)}\, d\phi \frac{8\pi}{M_P^2}\frac{V(\phi)}{\partial_{\phi}V(\phi)}},
\end{equation}
which for the quadratic potential becomes
\begin{equation}
a_{\text{SR}}(t)=a_0 e^{-\frac{2\pi}{M_P^2}\left[\phi_{\text{SR}}^2(t) - \phi_0^2\right]}.
\end{equation}

Lastly, we need the relation between cosmological and conformal time in order to write both $a(\eta)$ and $R(\eta)$. This relation can be obtained numerically from $\eta = \eta_0 + \int_{0}^t d t/a(t)$. These are the necessary ingredients for determining the frequency of the mode equation in this region, under the slow-roll approximation.

This regime is valid as long as the slow-roll parameter, $\epsilon_H=-{\dot{H}}/{H^2}$, is much smaller than one. When this no longer holds, at, say, $t>t_{*}$ with $t_{*}<0$, the equation of motion~\eqref{eq:InflatonEOMFlat} has to be solved numerically. The field begins to exit the inflationary regime and $t=\eta=0$ marks both the end of inflation and the beginning of reheating. At this point, the scale factor reaches the value $a_0$, which merely sets the scale and hence we take it to be $a_0=1$.

\subsection{Late reheating}
\label{subsec:latereheating}
For late times, well into the reheating epoch ($\eta_*\ll \eta \lesssim \eta_{\text{f}}$), and assuming $\eta_{\text{f}} < \eta_{\text{rh}}$\footnote{For sufficiently small masses, this is not the case. We will deal with this situation at the end of subsection~\ref{subsec:vacuumchoice}.}, where $\eta_{\text{rh}}$ denotes the end of reheating, one can find an approximate solution to \eqref{eq:InflatonEOMFlat} \cite{Cembranos2020}. We do not use it for obtaining our results, but it will be important for the discussion in subsection~\ref{subsec:vacuumchoice}. 
In this approximation, the Hubble rate reads
\begin{equation}
H(t)\simeq \frac{2}{3t}\left[1 - \frac{\sin{(2m_{\phi}t - 2\varphi)}}{2m_{\phi}t}+ \mathcal{O}(m_{\phi}^{-2}t^{-2})\right]^{-1},
\label{eq:HubbleRateInflaton}
\end{equation}
whereas the inflaton field is given by the expression
\begin{equation}
\phi = \frac{\Phi_0}{t}\sin{m_{\phi}t}\left[1 -\frac{\cos{2m_{\phi}t}}{2m_{\phi}t} + \mathcal{O}(m_{\phi}^{-2}t^{-2})\right],
\label{eq:LateReheatingInflaton}
\end{equation}
with $\Phi_0 \equiv {M_P}/{(\sqrt{3\pi}m_{\phi})}$.
This solution is valid as long as $m_{\phi}t \gg 1$, condition which is fulfilled during reheating, since, as we will see, the scale factor behaves as that of a matter dominated universe. Indeed, we can integrate $H(t)$ in order to approximately obtain the scale factor $a(t)$, 
\begin{equation}
a(t) = \mathcal{C} t^{2/3}\left[1 +  \mathcal{O}(m_{\phi}^{-2}t^{-2}) \right].
\end{equation}
The constant $\mathcal{C}$ is determined by requiring that the value of the scale factor at late times coincides with the one obtained from the numerical simulation in the previous region. One can now integrate the scale factor in order to obtain $t(\eta) = \left({\mathcal{C}}\eta/3\right)^3$.

Now that we have a solution for the inflaton field and the scale factor valid for late times, we can obtain the Ricci scalar from  \eqref{eq:RicciStressEnergy} by taking the solution for $\phi(t)$  to first order in $(m_{\phi}t)^{-1}$. We end up with
\begin{equation}
R = \frac{8}{3t^2}\left[2\sin^2{m_{\phi}t} - \left(\cos{m_{\phi}t} - \frac{\sin{m_{\phi}t}}{m_{\phi}t}\right)^2 + \mathcal{O}(m_{\phi}^{-3}t^{-3})\right].
\label{eq:LateRicci}
\end{equation}

With this, we are able to describe the frequency of the mode equation until very late times, for which the approximations derived in this subsection behave even better. The density of produced particles will be calculated at a sufficiently large time $\eta_{\text{f}}$, such that the particle production is negligible from that point in time onwards. 

\section{Particle production}
\label{sec:particleproduction}

Once we have determined the behavior of the background geometry during inflation and reheating, we can solve the mode equation in order to extract the Bogoliubov coefficients after the evolution.

\subsection{Solution to the mode equation}
\label{subsec:modeequation}

In order to compute the gravitational production once reheating has ended, we need to solve equation \eqref{eq:AuxiliaryModeEquation} from the onset of inflation at $t_i$ until a time $t_f$ well inside the adiabatic regime at the end of reheating, with the frequency of the oscillator determined by the background geometry described in the previous section. In a similar way as we did for the background dynamics in section \ref{sec:background}, the mode equation is solved in the regions
\begin{equation}
\eta = \begin{cases}
\eta_{\text{i}} \leq \eta \leq \eta_*, \quad \text{ Slow-roll approximation},\\
\eta_*\leq \eta \leq \eta_{\text{f}}, \quad \text{Numerical solution}.
\end{cases}
\end{equation}

Let us start with the slow-roll era. In a de Sitter geometry, the Hubble rate is exactly constant, $H_0$, the Ricci scalar is $R = 12 H_0$, and the scale factor reads $a(\eta) = 1/(1-H_0\eta)$. Therefore, the frequency \eqref{eq:Frequency} takes the form
\begin{equation}
    \omega_{k, \text{dS}}^2 = k^2 + \frac{\mu^2}{\left(\eta-\eta_0\right)^2}, \qquad \text{with} \quad \mu^2 = m^2/H_0^2 + 12(\xi-1/6)  ,
\end{equation}
where $H_0 = H(\eta_{\text{i}}) = 1/\eta_0$ is the Hubble rate at the beginning of inflation. The solution to equation \eqref{eq:AuxiliaryModeEquation} in this simplified scenario which asymptotically at $\eta \to -\infty$ behaves as a positive frequency plane wave is given by
\begin{equation}
v_{k, \text{dS}}(\eta) = \sqrt{\pi\abs{\eta - \eta_0}/2}\, e^{i\pi \nu   } H_{\nu}^{(1)}\left(k\abs{\eta - \eta_0}\right),  \qquad \nu = \sqrt{{1}/{4}- \mu^2}.
\label{eq:DeSitterSolution}
\end{equation}
This is the so-called Bunch-Davies solution \cite{Birrell1982}. Note that there is a critical value $\mu^2=1/4$ for which $\nu=0$, which separates the regimes of real and imaginary $\nu$. In particular, for~$m^2/H_0^2 \ll 1$, we can approximately write $\mu^2 \approx 12\left(\xi-1/6\right)$, and therefore $\mu^2 = 1/4$ for~$\xi = 3/16$. At this point, there is no gravitational pair production in a de Sitter geometry~\cite{Borrajo2020}, and this fact will be important for the analysis in section \ref{sec:particleproduction}. 

However, our background geometry is \emph{not} exactly de Sitter, but given by the inflaton dynamics derived in section \ref{sec:background}. Within the slow-roll approximation, valid from the start of inflation at $\eta_{\text{i}}$ until $\eta_*$, the mode equation to solve is
\begin{equation}
v^{\prime\prime}_k(\eta) + \omega_{k, \text{SR}}^2(\eta) v_k(\eta) = 0,
\label{eq:ModeEquationSR}
\end{equation}
where the scale factor and the Ricci scalar in $\omega_{k, \text{SR}}(\eta)$ correspond to the analysis in subsection~\ref{subsec:SlowRoll}. Nevertheless, in the slow-roll regime, and for a certain range in $k, m$, and $\xi$, we can approximate the solution satisfying Bunch-Davies initial conditions by (see subsection \ref{subsec:approximations} for details)
\begin{equation}
    v_{k, \text{SR}}(\eta) \simeq \sqrt{\pi\abs{\tau_k}/2} e^{i\pi\nu} H_{\nu}^{(1)}\left(k\abs{\tau_k}\right),\qquad 
\tau_k= \frac{\omega_{k, \text{SR}}(\eta)}{\omega_{k, \text{dS}}(\eta)}(\eta-\eta_{*, k}) + \eta_{*, k}-\eta_0,
\label{eq:ApproximateSRSolution}
\end{equation}
where $\eta_{*, k}$ marks the limit of validity of the approximation. From this point on, equation \eqref{eq:AuxiliaryModeEquation} has to be solved numerically, independently of the background dynamics being numerical or analytical, taking as initial condition solution \eqref{eq:ApproximateSRSolution} and its derivative at $\eta_{*, k}$. The frequency one has to use in this case is that in \eqref{eq:Frequency}.

\subsection{Choice of reference vacua}
\label{subsec:vacuumchoice}
The solution $v_k(\eta)$ to the mode equation is associated with a particular choice of vacuum: the one that behaves as a plane wave at $\eta \to -\infty$. The procedure in subsection \ref{subsec:modeequation} allows us to evaluate $v_k(\eta_{\text{f}})$. However, in order to obtain the Bogoliubov coefficient $\beta_k$, we also need~$u_k(\eta_{\text{f}})$, which is the solution to the mode equation associated with the vacuum at this point in time. Then, from the Bogoliubov coefficients $\alpha_k$ and $\beta_k$, we will be able to extract the number density of produced particles at $\eta_{\text{f}}$. This time is chosen such that particle production becomes negligible for later times, condition that is fulfilled in the adiabatic regime, i.e., when
\begin{equation}
\Bigg | \frac{\omega_k^{\prime}(\eta_{\text{f}})}{\omega_k^2(\eta_{\text{f}})} \Bigg| \ll 1.
\label{eq:AdiabaticCondition}
\end{equation}
The value of $\eta_{\text{f}}$ highly depends on the parameters of the scalar field, and in particular, it becomes larger as the mass $m$ decreases. This is why, for certain regions in parameter space, it may be convenient to use the late-time approximation for the background dynamics described in \ref{subsec:latereheating}, instead of solving numerically the equation of motion of the inflaton field. It is worth mentioning that at the same time, a smaller coupling $\xi$ to the curvature implies that the Ricci scalar oscillations, which are the main source of non-adiabaticity, are less important, therefore resulting in an earlier $\eta_{\text{f}}$ at which \eqref{eq:AdiabaticCondition} holds true.

As long as the background is not static, the meaning of vacuum will change in time. Nevertheless, if the evolution is adiabatic enough, namely condition \eqref{eq:AdiabaticCondition} is fulfilled, one can use the so-called adiabatic prescription to define the instantaneous vacuum at a given instant~$\eta_{\text{f}}$,
\begin{equation}
u_k(\eta_{\text{f}}) = \frac{1}{\sqrt{\omega_k(\eta_{\text{f}})}}, \qquad u_k^{\prime}(\eta_{\text{f}}) = -\frac{1}{\sqrt{\omega_k(\eta_{\text{f}})}}\left(i\omega_k(\eta_{\text{f}}) + \frac{1}{2}\frac{\omega_k^{\prime}(\eta_{\text{f}})}{\omega_k(\eta_{\text{f}})}\right).
\label{eq:AdiabaticVacuum}
\end{equation}
In fact, it is this feature that allows us to extrapolate the results obtained at $\eta_{\text{f}}$ to the present when considering fields that interact only gravitationally \cite{Ema2018, Cembranos2020}.

Introducing \eqref{eq:LateRicci} in \eqref{eq:Frequency}, one finds that, as long as $mt \gg 1$, particle production will be governed by the mass term of the frequency~\eqref{eq:Frequency}, namely
\begin{equation}
    \omega_k^2(\eta) \simeq k^2 + a^2(\eta)m^2.
\end{equation}
Since the scale factor at late times behaves as $a(\eta) \sim \eta^2$, condition \eqref{eq:AdiabaticCondition} is fulfilled soon after the Ricci scalar oscillations become unimportant. For masses of the order of the inflaton, this happens at a time $\eta_{\text{f}}$ small enough that we do not need to invoke the late-time solution for the background, since everything can be calculated numerically in an efficient way. This is not the case for masses smaller than the inflaton, for which production stabilizes after many, many oscillations, given that $mt \gg 1$ is fulfilled at later times. As a consequence, if we want to use the adiabatic vacuum description, we need to go up to a very large $\eta_{\text{f}}$, and therefore we need to use the analytic approximation for the inflaton dynamics described in~\eqref{eq:LateReheatingInflaton}.

Alternatively, we can take a different definition for the vacuum that allows us to calculate the number density of produced particles at $\bar{\eta}\ll\eta_{\text{f}}$, even for $m\ll m_{\phi}$. Although it will be still important in terms of adiabaticity, the oscillating term in \eqref{eq:Frequency} does not affect particle production at sufficiently large (numerically accessible)~$\bar{\eta}$, and therefore we can define the frequency
\begin{equation}
    \omega_k^{(\text{avg})\,2}(\eta)=k^2 + a^2(\eta)\left[m^2 + \left(\xi-1/6\right)\braket{R}(\eta)\right],
\end{equation}
where the Ricci scalar oscillations are averaged. We can take this frequency to calculate the \emph{averaged} vacuum
\begin{equation}
u_k^{(\text{avg})}(\bar{\eta}) = \frac{1}{\sqrt{\omega_k^{(\text{avg})}(\bar{\eta})}}, \quad u_k^{(\text{avg})\,\prime}(\bar{\eta}) = -\frac{1}{\sqrt{\omega_k^{(\text{avg})}(\eta)}}\left(i\omega_k^{(\text{avg})}(\bar{\eta}) + \frac{1}{2}\frac{\omega_k^{(\text{avg})\,\prime}(\bar{\eta})}{\omega_k^{(\text{avg})}(\bar{\eta})}\right).
\label{eq:AveragedVacuum}
\end{equation}
This prescription of vacuum is such that the spectrum of produced particles obtained at $\bar{\eta}$ essentially concides with the one given by the adiabatic vacuum at the time where we reach the adiabatic regime, $\eta_{\text{f}}$, namely
\begin{equation}
n_k^{(\text{avg})}\Big|_{\eta = \bar{\eta}} \simeq n_k^{(\text{ad})}\Big|_{\eta = \eta_{\text{f}}}.
\label{eq:DensityVacua}
\end{equation}
The larger discrepancies will reside in low wavenumbers, for which $k \sim a^2(\eta)\braket{R}$, but this region of momentum space is supressed in the total density of produced particles by a factor~$k^2$ (for details see next subsection), since
\begin{equation}
    n(m, \xi) = \int \frac{d^3\mathbf{k}}{(2\pi)^3} \bra{0}\hat{n}_k\ket{0}  = \int \frac{dk}{2\pi^2}k^2 \abs{\beta_k}^2.
\label{eq:DensityOfParticles}
\end{equation}
As a consequence, no differences are appreciated at the chosen $\bar{\eta}$. 

This procedure has a limitation: It is valid up to the smallest mass $m$ for which the dynamics presented here remain the same until $\eta_{\text{f}}$. If reheating ends before $\eta_{\text{f}}$ for a particular mass, in principle, the result provided by the averaged vacuum is not strictly correct. However, one can argue that production after reheating will be negligible when compared to the number density of particles that have already been produced. In fact, the dynamics of the Ricci scalar will be the same after the end of reheating, $\eta_{\text{rh}}$, since radiation does not contribute to the stress-energy tensor, and the scale factor will behave as $\eta$ instead of as $\eta^2$. Therefore, the comoving spectra obtained once the adiabaticity regime is reached can be regarded the same independently of $\eta_{\text{rh}}$ being before or after $\eta_{\text{f}}$. 

On the other hand, if this mechanism aims at explaining the observed abundance of dark matter, we have to require that production ends before the time when structures start to form, around $t \sim 10^{12}\, \text{s}$. If this is not the case, the dark matter abundance observed nowadays will not correspond to the one obtained in this analysis. Nevertheless, a simple estimation using the scale factor of a radiation-dominated universe shows that masses above the order of~$m\sim10^{-30} \, \text{eV}$ would reach adiabaticity early enough (i.e., the condition $mt \gg 1$ is fulfilled before $10^{12}\,\text{s}$). This is many orders of magnitude below the mass of fuzzy cold dark matter, and hence all the interesting range of masses lie within the regime of validity of our method.

\subsection{Slow-roll approximation for the solution to the mode equation}
\label{subsec:approximations}
During inflation, spacetime expands quasi-exponentially. More specifically, the number of $e$-folds
\begin{equation}
    \frac{a(t_0)}{a(t_i)} = e^{\mathcal{N}}
\end{equation}
is required to be such that $\mathcal{N} \approx 50-60$ \cite{Starobinsky1980, Guth1981, Linde1982}. Because eq.~\eqref{eq:AuxiliaryModeEquation} cannot be solved analytically, even considering a slowly rolling inflaton field, one would need to use numerical methods in order to find a solution. However, the large amount of $e$-folds to cover makes it more interesting and feasible to rely on an analytic approximation, such as~\eqref{eq:ApproximateSRSolution}. We dedicate this subsection to formally develop the approximation and to test its validity. For notational convenience, in the calculations that follow we will write $\eta-\eta_0$  as $\eta$, and drop the mode index $k$. Let us start by defining the following small parameters for given values of $k, m$ and~$\xi$ which will be useful in the following. 
\begin{itemize}
\item First, we have
\begin{equation}
    \epsilon(m, \xi) = \underset{ \eta\in I_1}{\text{max}}\Bigg|1-\frac{\omega_{{\text{SR}}}(  \eta; m, \xi)}{\omega_{{\text{dS}}}( \eta; m, \xi)}\Bigg|, \quad \text{with} \quad I_1 = (-\infty, \eta_1),
\end{equation}
where $\eta_1$ is chosen such that $\epsilon \ll 1$. Then, we can define $f(\eta; m, \xi)$ by 
\begin{equation}
    \frac{\omega_{\text{SR}}}{\omega_{\text{dS}}} = 1 + \epsilon f.
\end{equation}
By construction, $\abs{f(\eta)} \leq 1$ for $\eta \in I_1$. Moreover, $f^{\prime}(\eta) \geq 0$.

\item It will also be convenient to define
\begin{equation}
    \sigma(m, \xi) = \underset{\eta\in I_2}{\text{max}}\Big|f^{\prime}(\eta; m, \xi)\eta\Big|, \quad \text{with} \quad I_2 = (-\infty, \eta_2),
\end{equation}
and choose $\eta_2$ such that $\sigma \leq \epsilon$. Then, we introduce $g(\eta; m, \xi)$ as
\begin{equation}
    f^{\prime}(\eta) = \frac{\sigma g(\eta)}{\eta},
\end{equation}
for which again we have that $\abs{g_k(\eta)}\leq 1$ for $\eta \in I_2$.

\item Similarly, we define
\begin{equation}
    \rho(m, \xi) = \underset{\eta \in I_3}{\text{max}} \Bigg|\frac{\omega^{\prime}_{\text{dS}}(\eta)}{\omega_{\text{dS}}(\eta)}\eta \Bigg|, \quad \text{with} \quad I_3 = (-\infty, \eta_3),
\end{equation}
and choose $\eta_3$ such that $\rho \leq \epsilon$.

Now, we take $\eta_*=\text{min}(\eta_1, \eta_2, \eta_3)$ and $I=(-\infty, \eta_*)$, where $I$ is the interval for which the three parameters $\epsilon, \sigma, \rho$ are small. Note that $\eta_*<0$ since inflation ends at $\eta=0$. 

\item We also need $|\eta_*/\eta_0|>1$.
\end{itemize}

The task is to solve equation \eqref{eq:ModeEquationSR}, for which we define a new time coordinate $\zeta$ within the interval $I$,
\begin{equation}
    d\zeta = \frac{\omega_{\text{SR}}(\eta)}{\omega_{\text{dS}}(\eta)} d\eta= \left[1+\epsilon f(\eta)\right] d\eta.
\end{equation}
After integration until $\eta \in I$ and taking the absolute value, this becomes
\begin{equation}
    \abs{(\zeta - \zeta_*) - (\eta - \eta_*)} = \epsilon \Bigg|\int^{\eta_*}_{\eta}f(t)dt\Bigg| = \mathcal{O}(\epsilon)(\eta - \eta_*).
\end{equation}
Then, choosing $\zeta_*=\eta_*$, this can be expressed as
\begin{equation}
    \zeta = \eta\left[1 + \mathcal{O}(\epsilon)\right].
\end{equation}
We change time coordinates $\eta\to \zeta$ in the mode equation, which takes the form
\begin{equation}
    \ddot{w}( \zeta) + \omega_{\text{dS}}^2\left[\eta( \zeta)\right] w( \zeta ) + \epsilon f^{\prime}\left[\eta( \zeta)\right]\frac{\omega_{\text{dS}}^2\left[\eta(\zeta)\right]}{\omega_{\text{SR}}^2\left[\eta(\zeta)\right]} \dot{w}( \zeta)=0,
\label{eq:ModeEquationZeta}
\end{equation}
where $w( \zeta) = v\left[\eta( \zeta)\right]$ and the dot denotes here derivative with respect to $\zeta$. 

Let us analyze the last term. With this aim, we introduce the dimensionless time~$\bar \zeta=\zeta/\eta_0$. Then, in terms of $\bar \zeta$, the equation above has the same form except for the last term that acquires an extra factor. Using the definition of $f'$ and $\sigma$ above, the coefficient of this term is
\begin{equation}
\epsilon f'\frac{\omega_{\text{dS}}^2 }{\omega_{\text{SR}}^2 }\eta_0=
\epsilon\sigma g(1+\epsilon f)\frac{\eta_0}{\eta}=O(\epsilon^2)\frac{\eta_0}{\eta}
\end{equation}
If we choose $\eta_*$ such that $|\eta_*/\eta_0|>1$, as mentioned above, this coefficient is of order $O(\epsilon^2)$. Furthermore, the frequency in the second term of \eqref{eq:ModeEquationZeta} is
\begin{align}
    \omega_{\text{dS}}^2(\eta(\zeta)) &= \omega_{\text{dS}}^2\left( \zeta\left[1+\mathcal{O}(\epsilon)\right]\right)\\
    &= \omega_{\text{dS}}^2(\zeta)\left[1 + 2\frac{\omega_{\text{dS}}^{\prime}}{\omega_{\text{dS}}}\Bigg|_{\zeta}\cdot\zeta\,\mathcal{O}(\epsilon)\right] \\
    & = \omega_{\text{dS}}^2\left(\zeta\right)\left[1 + \mathcal{O}(\epsilon^2)\right],
\end{align}
provided that $|\zeta\,\omega_{\text{dS}}^{\prime}(\zeta)/\omega_{\text{dS}}(\zeta)| \leq \rho = \mathcal{O}(\epsilon)$. 
This is satisfied for $\zeta=\eta\left[1+O(\epsilon)\right]<\eta_*$, i.e., for $\eta<\eta_*$.
Thus, the equation for $w$ can finally be written as
\begin{equation}
    \ddot{w}(\zeta) + \omega_{\text{dS}}^2(\zeta) w(\zeta) = \mathcal{O}(\epsilon_k^2).
\end{equation}

We can perturbatively solve the differential equation by writting $w = w_{0} + \epsilon w_{1} + \mathcal{O}(\epsilon^2)$. The solution to order $\epsilon^0$ is nothing but the de Sitter modes \eqref{eq:DeSitterSolution},
\begin{equation}
    w_{0}(\zeta) = \sqrt{\pi\abs{\zeta}}\, e^{i\pi \nu} H_{\nu}^{(1)}\left(k\abs{\zeta}\right),  \qquad \nu = \sqrt{{1}/{4}- \mu^2},
\end{equation}
and as a consequence, $w_{k, 0}$ behaves asymptotically ($\zeta \to -\infty$) as a plane wave. On the other hand, the coefficients of the solution to order $\epsilon^1$ will satisfy the same original equation but with the initial conditions that $w_1(-\infty)=0$  and therefore $w_{1}$ is identically  zero. We can then write $w$ as
\begin{equation}
\begin{split}
    w(\zeta) &= w_{0}(\zeta )\left[1 + \mathcal{O}(\epsilon^2)\right] \\
    &=  \sqrt{\pi\abs{\zeta}}\, e^{i\pi \nu} H_{\nu}^{(1)}\left(k\abs{\zeta}\right)\left[1 + \mathcal{O}(\epsilon^2)\right].
\end{split}
\end{equation}

In order to undo the coordinate transformation $\zeta \to \eta$ while keeping the error up to~$\mathcal{O}(\epsilon^2)$, we need to consider the $\mathcal{O}(\epsilon^1)$ terms in $\zeta = \eta\left[1+\mathcal{O}(\epsilon)\right]$. For this, we note that
\begin{equation}
\begin{split}
    \Bigg| \left(\zeta - \eta_*\right) - \frac{\omega_{\text{SR}}(\eta)}{\omega_{\text{dS}}(\eta)} (\eta - \eta_*) \Bigg| &= \Bigg| \left(\eta - \eta_*\right) + \epsilon \int_{\eta_*}^{\eta} f(t) dt - \left[1 + \epsilon f(\eta)\right] \left(\eta - \eta_* \right) \Bigg|\\
    &=\epsilon\Bigg| \int_{\eta_*}^{\eta} f(t) dt - \int_{\eta^*}^{\eta} f(\eta) dt \Bigg| \\
    &\leq \epsilon \int_{\eta_*}^{\eta} \abs{f(t) - f(\eta)} dt \\
    &=\epsilon \int^{\eta}_{\eta_*} \left| f^{\prime}(\eta)(t - \eta) + \frac{1}{2!}f^{\prime\prime}(\eta)(t - \eta)^2 + \cdots\right|  dt \\
    &\leq \epsilon \left\{\left|\frac{1}{2} f^{\prime}(\eta)(\eta - \eta_*)^2\right| +\left| \frac{1}{3!}f^{\prime\prime}\left(\eta - \eta_*\right)^3 \right| + \cdots\right\}.
\end{split}
\end{equation}
This means that, as long as the  terms in curly brackets are of order $\mathcal{O}(\epsilon)$, we can write
\begin{equation}
    \zeta = \eta_* + \left[\frac{\omega_{\text{SR}}(\eta)}{\omega_{\text{dS}}(\eta)} + \mathcal{O}(\epsilon^2) \right] (\eta - \eta_*) = \eta_* + \frac{\omega_{\text{SR}}(\eta)}{\omega_{\text{dS}}(\eta)} (\eta - \eta_*)\left[1 + \mathcal{O}(\epsilon^2) \right].
\end{equation}
The first term is equal to
\begin{equation}
 \frac{1}{2}\sigma \left|g(\eta)\frac{\eta - \eta_*}{\eta} \right|=O(\epsilon).
\end{equation}
The next terms are of the form $f^{(n)}\left(\eta - \eta_*\right)^{n+1}/n!$, which numerically can be seen to be smaller than the first one.
 
\begin{figure}[t!]
\includegraphics[width=\textwidth]{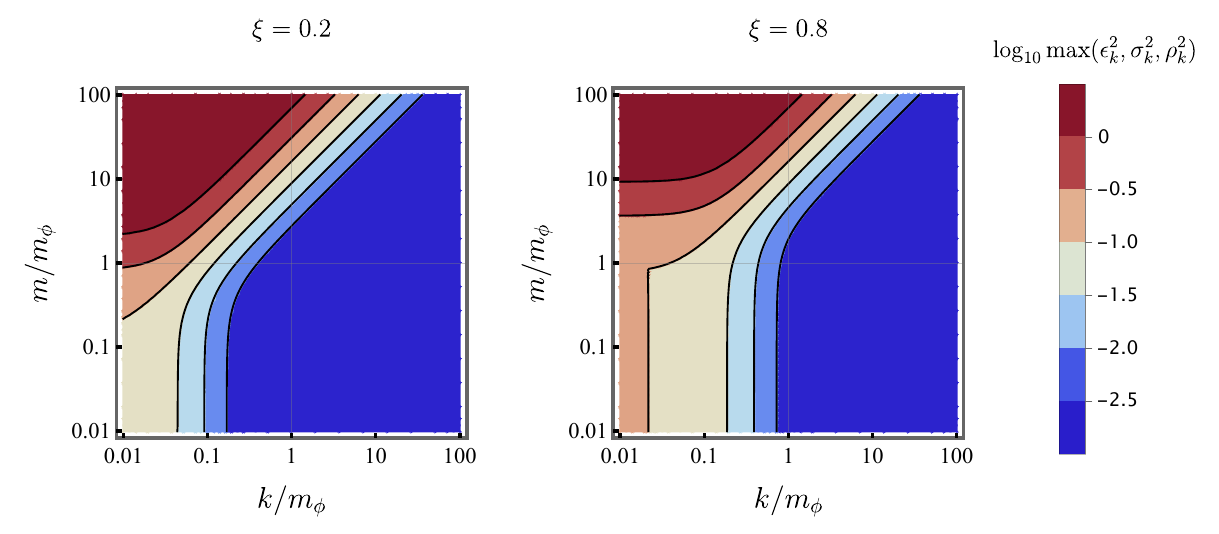}
\caption{Maximum of the errors squared as function of the wave number $k$ and the field mass $m$, for $\xi=0.2$ (left) and $\xi=0.8$ (right). We take $\eta_*=-500m_{\phi}$ for all values of $k$, $m$ and $\xi$.}
\label{fig:error}
\end{figure}
\begin{figure}[t!]
\includegraphics[width=\textwidth]{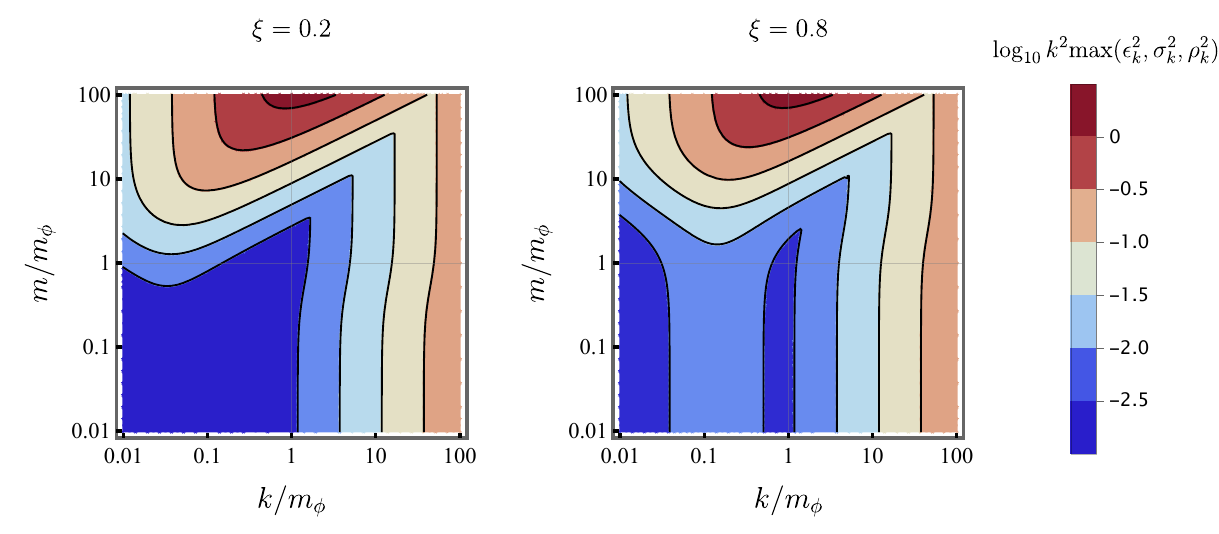}
\caption{Maximum of the errors squared times $k^2$ as function of the wave number $k$ and the field mass $m$, for $\xi=0.2$ (left) and $\xi=0.8$ (right). We take $\eta_*=-500m_{\phi}$ for all values of $k$, $m$ and $\xi$.}
\label{fig:k2error}
\end{figure}

Therefore, undoing the translation of $\eta$ to $\eta-\eta_0$ that we did at the beginning of this calculation, the solution to the mode equation can be written as \eqref{eq:ApproximateSRSolution} up to terms of order~$\mathcal{O}(\epsilon^2)$.
With fixed $\xi$, and choosing $\eta_*$ independent of $k$, the error $\epsilon_k$ increases with increasing $m$ and decreasing $k$.

When we numerically solve the mode equation \eqref{eq:AuxiliaryModeEquation} from $\eta_*$, the error in the initial condition coming from the slow-roll solution \eqref{eq:ApproximateSRSolution} carries through as
\begin{equation}
    v_k(\eta) = v_{k, 0}(\eta) \left[1 + \mathcal{O}(\epsilon_k^2)\right],
\end{equation}
such that $v_k(\eta) \to v_{k, \text{SR}}(\eta)$ as $\eta \to \eta_*$. Therefore, we have for the total density defined in~\eqref{eq:DensityOfParticles} that
\begin{equation}
    n(m, \xi) = \int_0^{\infty} \frac{dk}{2\pi^2} k^2 \abs{\beta_k}^2 = n_0 \left[1 + \frac{1}{n_0}\int_{0}^{\infty} \frac{dk}{2\pi^2}k^2\abs{\beta_{k, 0}}^2\mathcal{O}(\epsilon_k^2)\right],
\end{equation}
where $n_0 = \int_0^{\infty} \frac{dk}{2\pi^2} k^2 \abs{\beta_{k, 0}}^2$. Although the error $\epsilon_k$ increases as $k$ decreases, the factor $k^2$ compensates this increase for low $k$. Essentially, although $\epsilon_k^2$ increases for $k<m_{\phi}$, the quantity~$k^2 \epsilon_k^2$ remains small, whereas $\abs{\beta_{k, 0}}^2$ is roughly of the same order. More explicitly, for the calculations in this paper, we take $\eta_* = -500 m_{\phi}$, for which the maximum of the three small parameters squared, $\epsilon_k^2, \sigma_k^2, \rho_k^2$, as function of mass and wavenumber, for two different choices of coupling $\xi$, is shown in figure \ref{fig:error}. For $m\leq m_{\phi}$ and $k\geq 0.1m_{\phi}$, the error is of order~$\mathcal{O}(0.01)$ or smaller for the various values of $\xi$ considered, and thus the approximation is controlled in this regime. At the same time, we can observe in figure \ref{fig:k2error} that $k^2\epsilon_k^2$ decreases as we move to the low-part of the momentum range. This guarantees that this region of the spectrum is robust against errors in the mode equation approximation we used. 

On the other hand, from figure \ref{fig:k2error} we observe that the quantity $k^2\epsilon_k^2$ grows with $k$ for~$k>m_{\phi}$, since the decrease in $\epsilon_k^2$ (c.f. figure \ref{fig:error}) can not compensate the power $k^2$. However, gravitational production for high-momentum particles is very small, namely $\abs{\beta_k}^2 \approx 0$ for~$k\gg m_{\phi}$. As a consequence, $n(m, \xi) \approx n_0$ approximates well the total number density of particles produced, since the weight of wavenumbers $k\gg m_{\phi}$ is very small when compared to the rest of the spectrum.

\begin{figure}[t]
\centering
\includegraphics[width=\textwidth]{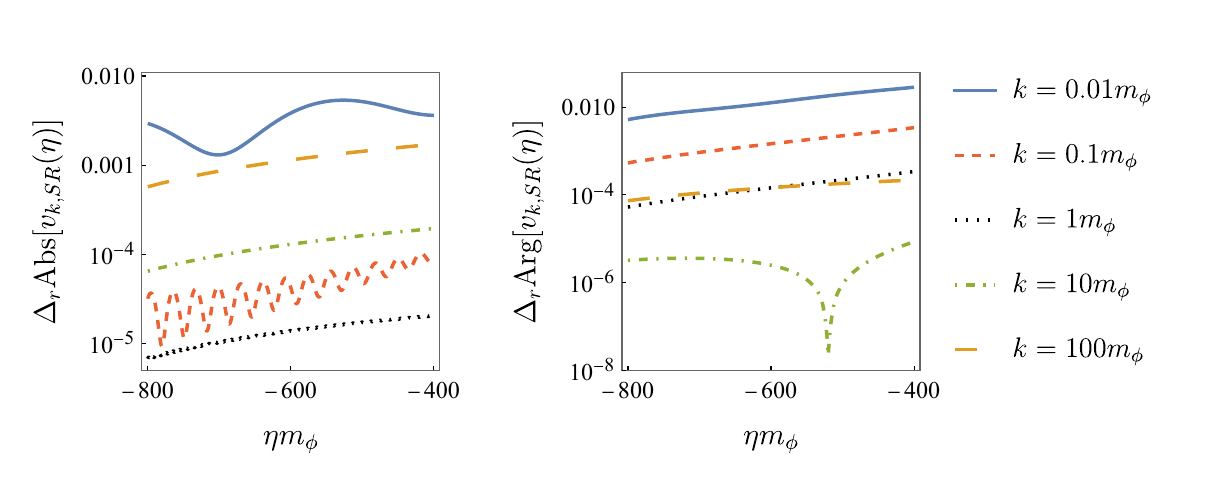}
\caption{Relative error in the absolute value (left panel) and the phase (right panel) of the numerical solution to the exact mode equation \eqref{eq:AuxiliaryModeEquation} compared to the analytical approximation~\eqref{eq:ApproximateSRSolution}, for wavenumbers ranging from $k=0.01m_{\phi}$ to $k=100 m_{\phi}$, and $m=m_{\phi}, \xi=0.5$. Here, we take~$\eta_{\text{dS}} = -1000/m_{\phi}$ and $\eta_* = -500/m_{\phi}$.}
\label{fig:SRvsWOSR}
\end{figure}

Furthermore, we can test the validity of \eqref{eq:ApproximateSRSolution} when compared to the numerical solution of \eqref{eq:AuxiliaryModeEquation} by putting ourselves in the following scenario: Let us assume that the geometry can be approximated by a de Sitter spacetime during the early stages of inflation, such that the solution \eqref{eq:DeSitterSolution} is valid for a region $\eta_{\text{i}}\leq\eta<\eta_{\text{dS}}$. At $\eta_{\text{dS}}$, slow-roll starts to matter, and deviations from the de Sitter solution $v_{k, \text{dS}}(\eta)$ occur. In this scenario, we explore two different paths to continue continuing solving the equation:
\begin{enumerate}
    \item We assume slow-roll inflation is a good description for the background dynamics in the region $\eta_{\text{dS}}\leq\eta<\eta_*$, and take as solution the approximation \eqref{eq:ApproximateSRSolution}.
    \item We solve numerically the exact equation of motion for the inflaton, eq. \eqref{eq:InflatonEOMFlat}, obtaining the frequency corresponding to \eqref{eq:AuxiliaryModeEquation}, equation which we again solve numerically. This solution, $v_k(\eta)$, will be valid even for $\eta \geq \eta_*$.
\end{enumerate}
In figure \ref{fig:SRvsWOSR}, we compare the analytical slow-roll solution with the exact numerical solution by plotting the relative difference between their absolute values, 
\begin{equation}
     \Delta_r \text{Abs}\left[v_{k, \text{SR}}(\eta)\right] \equiv \Bigg|\frac{\text{Abs}\left[v_k(\eta)\right] - \text{Abs}\left[v_{k, \text{SR}}(\eta)\right]}{\text{Abs}\left[v_k(\eta)\right]}\Bigg|,
\end{equation}
as well as their phase difference,
\begin{equation}
     \Delta_r \text{Arg}\left[v_{k, \text{SR}}(\eta)\right] \equiv \Bigg|\frac{\text{Arg}\left[v_k(\eta)\right] - \text{Arg}\left[v_{k, \text{SR}}(\eta)\right]}{\pi}\Bigg|.
\end{equation}
We do so for different wavenumbers, ranging from $k=0.01m_{\phi}$ to $k=100m_{\phi}$, denoted by the different shapes in figure \ref{fig:SRvsWOSR}. We have taken $\eta_{\text{dS}}=-1000/m_{\phi}$ as start of the slow-roll and~$\eta_* = -500/m_{\phi}$ as the time when the slow-roll approximation breaks down. For $k=m_{\phi}$, the relative error is very small, of order $\sim 10^{-4}$ at $\eta_*$. For wavenumbers larger than the mass of the inflaton, $k>m_{\phi}$, the approximation is still good, although it worsens. On the other hand, the error for $k=0.01 m_{\phi}$ starts becoming significant, and gets worse for $k<0.01 m_{\phi}$. However, the corresponding region of the spectrum of produced particles is highly suppressed, as discussed above, and therefore the contribution to the total density of particles is negligible. Similarly, particle production is very small for wavenumbers larger than $k>100m_{\phi}$, and therefore the range of interest in $k$ is under control. Hence, we can assume the approximation is valid in the region $\eta_{\text{dS}}\leq\eta<\eta_*$.

Note that if this solution behaves well in this region, it has to become an even better approximation before $\eta_{\text{dS}}$, since the further towards the past we go, the more de Sitter-like is the geometry. Thus, eq. \eqref{eq:ApproximateSRSolution} can be taken as well as a solution to the mode equation in the region $\eta_{\text{i}}\leq \eta <\eta_{\text{dS}}$.  Under this approximations, eq. \eqref{eq:AuxiliaryModeEquation} can be solved analytically from the start of inflation, $\eta_{\text{i}}$, until $\eta_*$, for which the slow-roll approximation starts to fail. From there, the mode equation is solved numerically.

\subsection{Adiabaticity and oscillations}
\label{subsec:adiabaticity}

\begin{figure}[t!]
    \centering
    \includegraphics[width=0.5\textwidth]{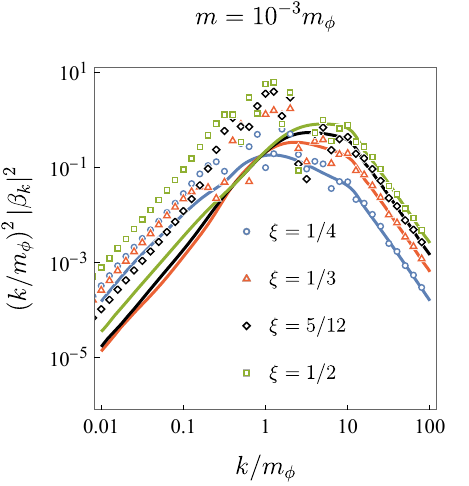}
    \caption{Spectra of produced particles of mass $m=10^{-3}m_{\phi}$ and different values of $\xi$, obtained with the adiabatic prescription of the vacuum. The dots correspond to $\eta = 16.33/m_{\phi}$, before the adiabatic regime has been reached for this value of the mass. The solid lines correspond to $\eta = \eta_{\text{f}} = 100/m_{\phi}$, when most of the particles have been produced.} 
    \label{fig:AdiabaticityLog}
\end{figure}

\begin{figure}[t!]
    \centering
    \includegraphics[width=0.5\textwidth]{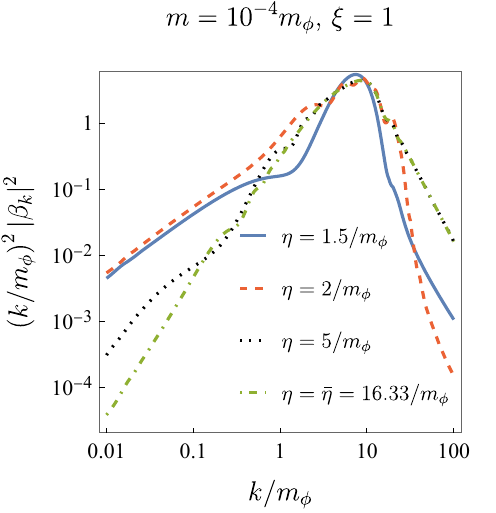}
    \caption{Spectra for $m=10^{-4}m_{\phi}$ and $\xi=1$, obtained with the averaged vacuum prescription, for different instants of time. The spectrum stabilises after very many oscillations of the curvature scalar.} 
    \label{fig:OscillationsInfluence}
\end{figure}

In order to illustrate the importance of the choice of vacuum, we studied the evolution of spectra when calculated using prescription \eqref{eq:AdiabaticVacuum} before the dynamics has entered the adiabatic regime. As an example, we plotted in figure \ref{fig:AdiabaticityLog} the spectra of particles with mass $m = 10^{-3}m_{\phi}$ obtained at two different times. The dots correspond to $\eta = 16.33/m_{\phi}$, whereas the solid lines denote $\eta = \eta_{\text{f}} = 100/m_{\phi}$. For this particular choice of mass, the latter time lies within the adiabatic regime, and this is the reason why the non-adiabatic dots relax to their final value as we approach this limit. As expected, the effect is less noticeable the lower the coupling to the geometry is, as it is the main source of non-adiabaticity in the frequency.

At the same time, we also characterized the importance of the first oscillations of the curvature scalar in the final spectrum of produced particles, obtained with the averaged vacuum defined in eq. \eqref{eq:AveragedVacuum}. As can be seen in figure \ref{fig:OscillationsInfluence}, even after several oscillations of~$R(\eta)$ (for example, at $\eta = 2/m_{\phi}$), the production changes greatly if one compares with the obtained spectra at $\bar{\eta}$. Even when looking only at the total number of produced particles in eq.~\eqref{eq:DensityOfParticles}, differences are still significant. We observe that the spectrum does not stabilize until $\eta \simeq 5/m_{\phi}$, which for our model means after hundreds of oscillations of the curvature scalar $R(\eta)$. With this, we want to stress that obtaning the particle production after one or two oscillations does not account for the whole process.

\section{Spectra of particles and total density}
\label{sec:spectra}

Let us finally give the results for the spectra of produced particles as function of the parameters of the field, the mass $m$, and the coupling to the curvature $\xi$. The following calculations have been performed using the averaged vacuum prescription at $\bar{\eta}=16.33/m_{\phi}$. 

\begin{figure}[t!]
    \centering
    \includegraphics[width=0.5\textwidth]{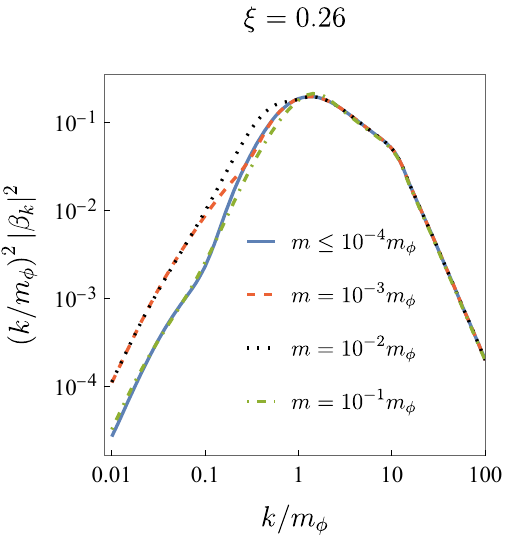}
    \caption{Spectrum of particles for masses below the mass of the inflaton, with $\xi=0.26$. For very small masses ($m\leq 10^{-4}m_{\phi}$), production is dominated by curvature. In the region $10^{-3}\leq m \leq 10^{-1}m_{\phi}$, differences in production due to the mass can be noticed, especially for low values of $k/m_{\phi} \simeq 0.1-1$.} 
    \label{fig:SpectrumMassesBelow}
\end{figure}

\begin{figure}[t!]
    \centering
    \includegraphics[width=0.5\textwidth]{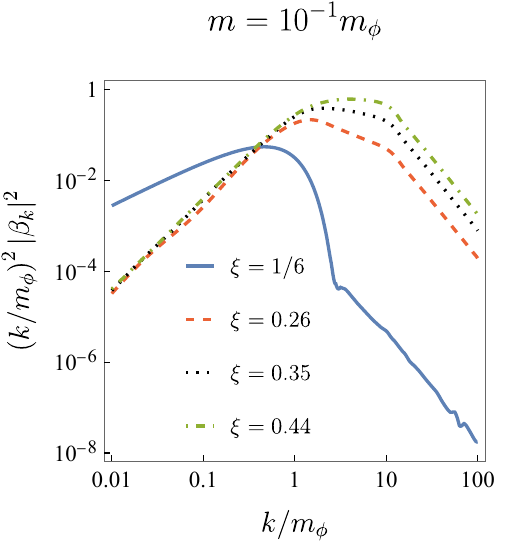}
    \caption{Spectra for $m=10^{-1}m_{\phi}$ and several values of the coupling $\xi$. Particle production increases when the curvature term becomes more important, and the maximum of the spectrum is shifted towards higher values of $k$.} 
    \label{fig:Spectrum01Mphi}
\end{figure}

We explore first the regime of masses below the inflaton mass. Represented by the solid line in figure \ref{fig:SpectrumMassesBelow}, we have masses $m\leq 10^{-4}m_{\phi}$. For these values, the mass contribution to the frequency becomes negligible, and the dynamics is entirely given by the coupling to the geometry. The spectra lie on top of each other, with very small differences in the low values of $k \sim a(\eta)m$. We observe, however, slight differences in the shape of the spectrum when increasing the mass, especially for small wavenumbers, as the rest of the curves in figure~\ref{fig:SpectrumMassesBelow} show. We can choose a mass in this regime, $m=10^{-1}m_{\phi}$, and explore the influence of the coupling $\xi$ in the final result. This is shown in figure \ref{fig:Spectrum01Mphi}, where one observes increasing production of particles with larger values of the coupling.

\begin{figure}[t!]
    \centering
    \includegraphics[width=0.5\textwidth]{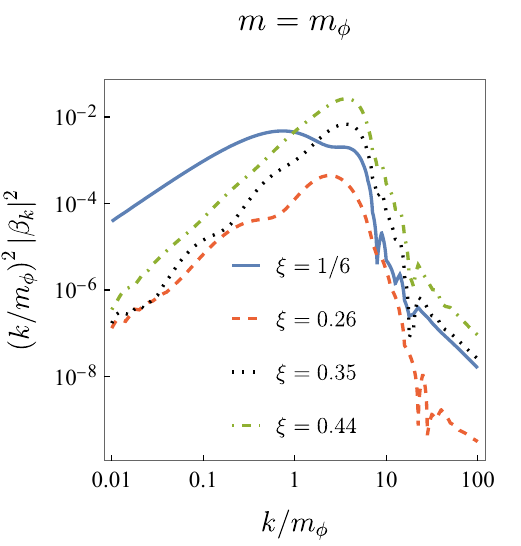}
    \caption{Spectra of particles with the mass of the inflaton, for different values of the coupling. In this particular case, increasing the coupling does not translate directly into an increase of particle production. This can be more clearly seen by examining the total density of particles.} 
    \label{fig:SpectrumInflatonMass}
\end{figure}

Lastly, let us come to the mass of the inflaton, whose corresponding spectra are shown in figure \ref{fig:SpectrumInflatonMass}. In such a case, it is harder to characterize the behavior with $\xi$. It is clear, nevertheless, that particle production decreases as the mass of particles becomes larger.

\begin{figure}[t!]
    \centering
    \includegraphics[width=\textwidth]{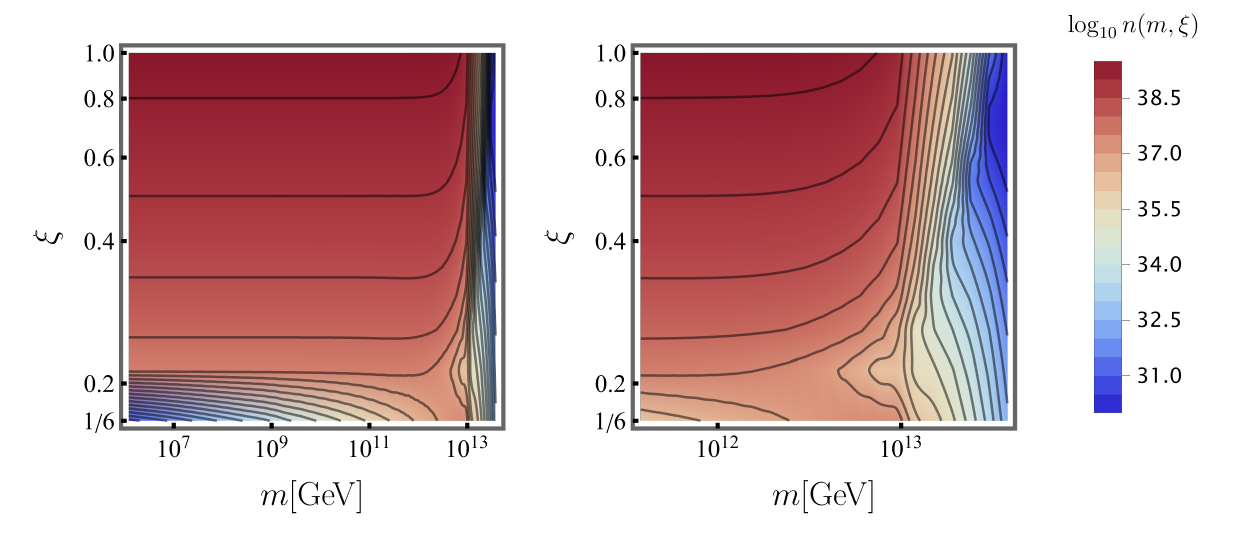}
    \caption{Logarithm of the total density of produced particles for different values of $m$ and $\xi$. In order to give the mass and density in units of GeV, we took $m_{\phi}=1.2 \times 10^{13} \, \text{GeV}$ for the mass of the inflaton. We explore a wide range of masses in the left panel while we focus on a smaller region close to the mass of the inflaton on the right panel in order to appreciate the dependence of the total density with the coupling $\xi$.} 
    \label{fig:TotalDensity}
\end{figure}

It is easier to characterize particle production in this regime using the total number density of particles \eqref{eq:DensityOfParticles}, which we show in figure \ref{fig:TotalDensity} as function of the two parameters of the field, $m$ and $\xi$. Here, one clearly sees that the prediction is independent of the value of the mass as long as it is below $m \sim 10^{-2}m_{\phi}$, in particular for a sufficiently high value of the coupling, $\xi \gtrsim 0.2$. In this case, the mass is completely negligible when compared to the dynamics of the curvature scalar. Only when the coupling to the curvature is close to $\xi \sim 1/6$, the production of particles is still sensible to $m$, up to $m \sim 10^{-7}m_{\phi}$. For this value, even in the conformal case, the relevant wavenumbers, $k \sim a(\eta)m$, are too suppressed to make a difference. In all these regime of low masses, the number of produced particles increases with larger coupling $\xi$. Closer to the mass of the inflaton, $10^{-2}m_{\phi}<m<m_{\phi}$, the fact that a heavier particle translates into a lower production becomes apparent. Lastly, in the region around the mass of the inflaton, $m\sim m_{\phi}$, the behavior with the coupling is different, and production may even decrease when raising the value of $\xi$. In fact, there appears to exist a critical value $\xi_{\text{c}} \simeq 0.22$ which separates two qualitatively different regimes. 
As we commented previously, this value is related to the parameter $\mu^2=1/4$ of the Hankel functions, which were a good approximation of the mode functions of our problem. 
\begin{figure}[t!]
    \centering
    \includegraphics[width=\textwidth]{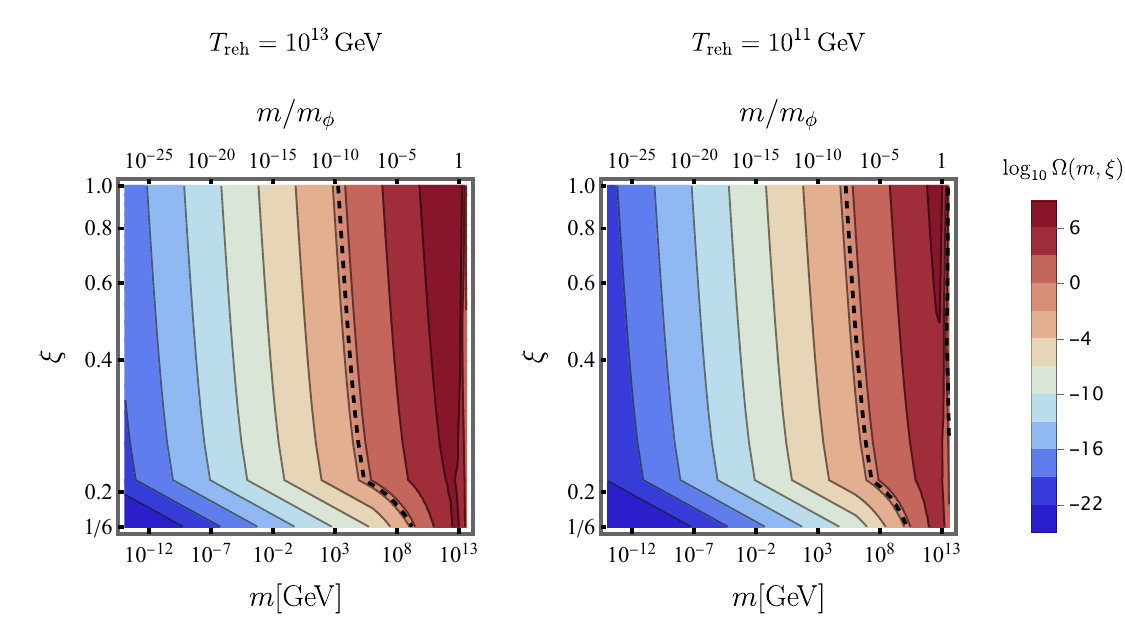}
    \caption{Logarithm of the predicted abundance of dark matter today for different values of $m$ and~$\xi$, and a reheating temperature of $T_{\text{reh}}=10^{13} \, \text{GeV}$ (left) and $T_{\text{reh}}=10^{11} \, \text{GeV}$ (right). The observed abundance corresponds to the dashed line. In order to give the mass and density in units of GeV, we took $m_{\phi}=1.2 \times 10^{13} \, \text{GeV}$ for the mass of the inflaton.} 
    \label{fig:Abundance}
\end{figure}For $m<m_{\phi}$, the number density drops very rapidly if $\xi<\xi_{\text{c}}$. For $m\sim m_{\phi}$, $\xi_{\text{c}}$ is the value below which production decreases with $\xi$, and above which it increases. This is also illustrated in figure \ref{fig:SpectrumInflatonMass}, where production for $\xi=1/6$ is larger than for $\xi=0.26$, and from there it increases again with the coupling. Moreover, we observe the expected strong suppression in the number density of produced particles for masses above the mass of the inflaton. We can confirm this behavior by calculating the spectra for even higher masses, provided we select a negative enough $\eta_*$ --- and therefore leading to a longer computation --- in this case, as explained in \ref{subsec:approximations}. Note that we took $m_{\phi}=1.2 \times 10^{13} \, \text{GeV}$ for the mass of the inflaton, and as a consequence, the density in figure \ref{fig:TotalDensity} is given in units of $\text{GeV}^{3}$.

Finally, one can consider these gravitationally produced scalar particles as dark matter. In this case, it is  necessary to compare the resulting abundance with observations. The physical density of produced particles is related to the comoving density shown in figure \ref{fig:TotalDensity} only by the scale factor. Assuming that the scalar field is non-interacting, which is mandatory for the gravitational production to be important, as it cannot reach thermal equilibrium, the evolution of the density of created particles from $\eta_{\text{rh}}$ until today will be dictated solely by the dilution due to the isentropic expansion of the background. The predicted abundance can be written in terms of the background radiation temperature \cite{Cembranos2020} as
\begin{equation}
    \Omega (m, \xi) = \frac{8\pi}{3M_P^2H^2_{\text{today}}}\frac{g_{*S}^{\text{today}}}{g_{*S}^{\text{rh}}}\left(\frac{T_{\text{today}}}{T_{\text{rh}}}\right)^3 m \, \frac{n(m, \xi)}{a_{\text{rh}}^3},
\end{equation}
where $T_{\text{today}}$ and $T_{\text{rh}}$ are the radiation temperature today and at the end of reheating, respectively, and $g^{\text{today}}_{*S}$ and $g^{\text{rh}}_{*S}$ are the corresponding relativistic degrees of freedom. The scale factor at the end of reheating, $a_{\text{rh}}$ is obtained using that, when radiation dominates, at~$\eta_{\text{rh}}$, the Hubble rate can be written as
\begin{equation}
    H^2_{\text{rh}} = \frac{8\pi}{3M_P^2}\frac{\pi^2}{30}g_{*S}^{\text{rh}}T_{\text{rh}}^4,
\end{equation}
which allows one to obtain $\eta_{\text{rh}}$ as function of the reheating temperature. This sets an upper limit on the reheating temperature, since $\bar{\eta} < \eta_{\text{rh}}$ for $T_{\text{rh}} \lesssim 10^{13} \, \text{GeV}$. Let us remark that for the region of parameter space considered, the comoving energy density of the spectator field, namely $mn(m, \xi)$, is many orders of magnitude lower than that of the inflaton, and therefore neglecting backreaction is a well justified assumption.

The abundance is represented in figure \ref{fig:Abundance} for different reheating temperatures, together with the observed dark matter abundance, given by the dashed line. We observe that the proposed mechanism can explain observations if the dark matter candidate is light enough ($m \leq 10^8 \, \text{GeV}$ for $T_{\text{rh}}=10^{13}\, \text{GeV}$), independently of the value of the coupling $\xi$ for the range that we considered. In addition, heavier particles can also reach the observed dark matter abundance since their production is strongly suppressed above the inflaton mass. 

\section{Conclusions}
\label{sec:conslusions}

Gravitational particle production is a very interesting process due to its universality. It only requires the studied field to interact with gravity. Even without a direct coupling to the inflaton, as it is the case of spectator fields such as the one we have studied, it can give rise to a significant abundance for the species considered after the heavy expansion of spacetime during the early stages of the universe. However, predictions need for a definition of vacuum after reheating, since the non-static geometry leads to certain ambiguity in the meaning of \emph{particle}.

In this manuscript, we studied the production of massive, scalar particles whose dynamics is described by a non-minimally coupled to gravity action. However, the discussion on the validity of the definition of vacuum is pertinent when considering any other field as well. First, we have provided a method for solving in a complete form the background dynamics, governed by a single scalar inflaton field. For this, we did not have to assume a de Sitter geometry of spacetime, which would significantly change the amount of particles produced. Although we make a choice of potential, this procedure can be extended to other cases as well. We provided an analytic approximation to the solution of the slow-roll mode equation where the error is well under control in our parameter region of interest. More importantly, we showed that, for masses smaller than the inflaton mass, the commonly used adiabatic prescription for the vacuum determines correctly the production of particles after reheating only when calculated at very late times. Moreover, we define an alternative vacuum choice that allows one to obtain the right abundance when calculating particle production at a much earlier time. This allowed us to explore the contribution of the first oscillations to the total number of produced particles, revealing that the spectra only stabilizes after hundreds of periods. Lastly, after all these considerations have been taken into account, we analyzed both the spectra and the total density of particles for different values of the mass of the field and its coupling to the curvature scalar. When regarded as dark matter, the production of the spectator field can be directly related to the abundance that would be observed today if one assumes no couplings to any other fields also after reheating. In particular, we find agreement with the observed dark matter abundance for a certain range of masses and couplings of the spectator field. Moreover, this analysis can be used to constrain the values of the field parameters by demanding that the predicted dark matter abundance does not exceed observations.


\section*{Acknowledgements}

This work was partially supported by the MICINN (Ministerio de Ciencia e Innovación, Spain) projects PID2019-107394GB-I00/AEI/10.13039/501100011033 (AEI/FEDER, UE), PID2020-118159GBC44, and PID2022-139841NB-I00, COST (European Cooperation in Science and Technology) Actions CA21106 and CA21136. Additionally, Á.P.-L. is supported by the MIU (Ministerio de Universidades, Spain) fellowship FPU20/05603. JARC acknowledges support by Institut Pascal at Université Paris-Saclay during the Paris-Saclay Astroparticle Symposium 2022, with the support of the P2IO Laboratory of Excellence (program “Investissements d’avenir” ANR-11-IDEX-0003-01 Paris-Saclay and ANR-10-LABX-0038), the P2I axis of the Graduate School of Physics of Université Paris-Saclay, as well as IJCLab, CEA, APPEC, IAS, OSUPS, and the IN2P3 master projet UCMN. Finally, JMSV acknowledges the support of the Spanish Agencia Estatal de Investigaci\'on through the grant “IFT Centro de Excelencia Severo Ochoa CEX2020-001007-S".

\appendix

\section{Parameters}
\label{app:parameters}

In the majority of the analyses, we have left all the quantities expressed in terms of the mass of the inflaton, $m_{\phi}$, which sets up the scale of the problem. When it has been necessary to assume a numerical value for such a mass, we have taken $m_{\phi} = 1.2 \times 10^{13} \, \text{GeV}$. Accordingly, the Planck mass $M_P$ has the value $M_P = 1.02 \times 10^6 m_{\phi}$. 

The initial value for the inflaton field, under the slow-roll assumption, is taken to be~$\phi_{\text{SR}}(t_i) = \phi_i = 3M_P$. When inflation ends, at $t = 0$, the field value is $\phi_{\text{SR}}(t=0)= \phi_0 = 0.5M_P$. The slow-roll approximation can then be used to extract $t_i \simeq -15.35/m_{\phi}$ as the time when inflation starts. Equation of motion \eqref{eq:InflatonEOMFlat} can also be solved numerically taking as initial conditions the same as for slow-roll, $\phi(t_i)=\phi_i$, and the derivative of the approximate solution at this point, $\phi^{\prime}(t_i)=\phi_{\text{SR}}^{\prime}(t_i)$. Both solutions will be very close up to $t_*$, where the slow-roll approximation starts to break down. Then, $\phi(t=0)$ slightly deviates from $\phi_0$. The scale factor is chosen such that $a(t=0)=a_0=1$. Slow-roll is a assumed to be a good approximation until $\eta_*=-500/m_{\phi}$.

Unless the contrary is expressly stated, particle production is calculated using the averaged vacuum prescription at $\bar{\eta} = 16.33/m_{\phi}$. The range of masses explored is $10^{-7}m_{\phi}\leq m \leq 10^{0.5}m_{\phi}$, although for obtaining figure \ref{fig:Abundance} it is assumed that production is the same for~$m\leq 10^{-7}m_{\phi}$. On the other hand, the coupling $\xi$ is such that $1/6 \leq \xi \leq 1$.


\bibliographystyle{JHEP.bst}
\bibliography{references.bib}

\end{document}